\documentstyle[preprint,aps,prd]{revtex}

\newcommand{\be}{\begin{equation}}
\newcommand{\ee}{\end{equation}}
\newcommand{\bea}{\begin{eqnarray}}
\newcommand{\eea}{\end{eqnarray}}

\begin{document}

\title{Chaos and Semiclassical Limit in Quantum Cosmology}

\author{Esteban Calzetta \thanks{EMail: calzetta@iafe.uba.ar}}

\address{Instituto de Astronomia y F\'{\i}sica del Espacio,\\ Casilla
de Correo 67, Sucursal 28, 1428 Buenos Aires, Argentina \\ and \\
Departamento de F\'{\i}sica, Facultad de Ciencias Exactas y Naturales,
Universidad de Buenos Aires,\\ Ciudad Universitaria, 1428 Buenos Aires,
Argentina}

\author{Juan Jos\'e Gonzalez \thanks{EMail: gonzalez@dfuba.df.uba.ar}}

\address{Departamento de F\'{\i}sica, Facultad de Ciencias Exactas y
Naturales,Universidad de Buenos Aires,\\ Ciudad Universitaria, 1428
Buenos Aires, Argentina}

\date{\today}

\maketitle

\begin{abstract}

     In this paper we present a Friedmann-Robertson-Walker cosmological
     model conformally coupled to a massive scalar field where
     the WKB approximation fails to reproduce the exact solution to the
     Wheeler-DeWitt equation for large Universes. The breakdown of the WKB
     approximation follows the same pattern than in semiclassical physics
     of chaotic systems, and it is associated to the development of small
     scale structure in the wave function. This result puts in doubt the
     ``WKB interpretation'' of Quantum Cosmology.

\end{abstract}


\narrowtext

\section{Introduction}

    Quantum Cosmology \cite{Halliwell90} tries to provide a complete
    description of the
    universe. Since the present universe obeys classical laws with great
    precision, an acceptable Quantum Cosmology must predict a ``quantum
    to classical'' transition at a certain time in the cosmic evolution.
    In the same manner, it must provide an interpretation of the initial
    conditions of the classical evolution, in terms of quantum processes.
    For these reasons, it is clear that the problem of the correspondence
    between classical and quantum behavior is a central point in the
    development of quantum cosmology.

    To explain our observations it is enough to interpret the wave function
    of the universe in the semiclassical limit \cite{Kuchar92}. Moreover,
    it has been argued that it is sufficient to have a Wentzel-Kramers-
    Brillouin (WKB) state or a
    Gaussian state to have a valid  semiclassical description
    \cite{Vilenkin89}. This is the basis of the so-called ``WKB
    interpretation of Quantum Cosmology''. According to this interpretation,
    we can understand the semiclassical limit as the limit $m_p \to \infty$,
    where $m_p$ stands for the Plank mass. In this limit the
    wave function of the universe, may be understood as a superposition of
    solutions of the WKB type \cite{Hartle86},

    \be \Psi = \sum_ \alpha C_ \alpha \, \exp [{i \over \hbar}
    S_\alpha ] \ee

    \noindent where the $S_\alpha$ are solutions to the classical Hamilton
    Jacobi equation and the $|C_ \alpha |^2$ may be interpreted as the
    relative probability the $\alpha$
    classical solution. However, we want to obtain the semiclassical
    limit when the universe has expanded beyond a certain size, and not in
    the limit $m_p \to \infty$. The $m_p \to \infty$ limit of the wave
    function of the Universe would provide nevertheless information on
    the large size behavior, if the limits $m_p \to \infty$ and
    $a \to \infty$ commuted, where $a$
    stands for the radius of the universe in a Robertson-Walker model.
    Although more recent papers have suggested the necessity of
    including  decoherence to the WKB limit, in order to obtain a truly
    semiclassical state \cite{PazSinha91}, it is commonly assumed in the
    literature on quantum cosmology that the wave function of the universe
    adopts a WKB form at some point in the cosmic evolution, and retains
    it for ever after.

    This problem is similar to the problem of the commutativity of the
    limits $\hbar \to 0$ and $t \to \infty$ in ordinary quantum
    mechanics. The semiclassical wave function has been studied in great
    detail, and it has been shown that for
    irregular systems (chaotic systems) there exists a breakdown of the
    validity of the WKB approximation \cite{Berry83,Ozorio90}. The existence of
    chaotic cosmologies \cite{CalzettaElHasi93} suggests that a
    similar breakdown in the WKB approximation might occur in
    cosmological models. This would put in doubt the WKB interpretation
    of quantum cosmology.

    However, the problem of quantum cosmology cannot be referred directly
    to the quantum mechanical problem, because the Wheeler-DeWitt equation,
    the equation that governs quantum cosmology, is of second order. Therefore,
    we need to analyze the problem from first principles; it is convenient
    to begin such analysis with a simple, exactly solvable model.

    In this paper, we present such a model, namely a spatially flat
    Friedmann-Robertson-Walker (FRW) model
    coupled to a scalar field. Choosing a suitable
    coupling of the field to the radius $a$ of the universe, we solve
    the Wheeler-DeWitt equation exactly. For the same model we calculate
    the semiclassical wave function, and we compare the evolution of this
    wave function with the exact one, checking the validity of the WKB
    approximation.

    The results obtained in this paper confirm the suspicion that the WKB
    approximation breaks down in quantum cosmology for large universes.
    This puts in doubt the WKB interpretation in its original formulation.
    The WKB approximation may still be valid, however, in more realistic
    models, for example, those including decoherence.

\section{The Model}

    Our cosmological model assumes a FRW spatially flat geometry,
    conformally coupled to a real massive scalar field $\Phi$. We shall
    suppose a $\Phi(r,\eta)=exp(i \, k.r) \phi (\eta)/ a(\eta)$,
    with $k<<1$. This is a dynamical system
    with two degrees of freedom, $a$ and  $\phi$, and Hamiltonian
    \cite{CalzettaElHasi93}

    \be H={1\over  2} \{{-{P_a}^2\over  {m_p}^2}+{\pi_\phi}^2+
    [k^2+m^2 V(a)] \phi^2 \}=0 \ee

    Where $\pi_\phi$ y $P_a$ are the momenta conjugated to $\phi$ and
    $a$ respectively.

    For the numerical calculation we choose the potential

    \be V(a)=\sum_{n=1}^\infty (2n-1) \, \Theta (|a|-\delta a n)
    \delta a^2 \label{v} \ee

    where $\Theta(x)$ is the Heaviside step function and $\delta a$ is
    the step length (fig.\ref{Fig1}).

    Introducing dimensionless variables $P_a/m_p \, k^2 \to P_a$, $a \,
    m_p / k \to a$, $\pi_\phi / k^2 \to \pi_\phi$, $\phi / k \to \phi$,
    $m / m_p \to m$, the adimensional Hamiltonian reads

    \be H={1\over  2}\{-{P_a}^2+{\pi_\phi}^2+[1+m^2\sum_{n=1}^\infty (2n-1)
    \, \Theta(|a|-\delta a n)\delta a^2]\phi^2 \}=0 \ee

    We can reduce the system's freedoms by one, using the Hamiltonian
    constraint. Hamilton's equations for the reduced system in the
    range  $n \delta a < a \leq (n+1) \delta a$, are

    \be {d\phi\over  da}={\delta h\over \delta\pi}={\pi\over \sqrt{\pi^2+
    {\omega_n}^2 \phi^2}}={\pi\over  E_n} \ee

    \be {d\pi\over  da}=-{\delta h\over  \delta\phi}={-{\omega_n}^2\phi\over
    \sqrt{\pi^2+{\omega_n}^2\phi^2}}={{-{\omega_n}^2\phi}\over E_n} \ee

    Where now $a$ plays the role of time, $h=-P_a=E_n
    = \sqrt{\pi^2+{\omega_n}^2\phi^2}$ and

    \be \omega_n(a)=[1+m^2 n^2 \delta a^2]^{1/2} \label{c} \ee

    is the frequency in each step. These
    equations of motion correspond to an harmonic oscillator of frequency
    ${\omega_n \over E_n}$, so the solution for the reduced system is

    \be \phi^{(n)}(\delta a)=\phi^{n+1}=\phi^n\cos({\omega_n\over  E^n}
    \delta a)+{{\pi_\phi}^n\over \omega_n}\sin({\omega_n\over  E^n}\delta
    a) \label{a}\ee

    \be \pi^{(n)}(\delta a)=\pi^{n+1}=-\omega_n\phi^n\sin({\omega_n\over
    E^n} \delta a)+\pi^n\cos({\omega_n\over  E^n}\delta a) \label{b} \ee

    These equations define a stroboscopic map of period $\delta a$, where
    $\phi^n$ and $\pi^n$ are the values of the field and its conjugated
    momentum in the border of the nth step.

    As we cross the threshold from one step to the next, the energy
    changes from $E_n$ to $E_{n+1}$.
    This non-conservation of the energy is a signal of the
    non-integrability of this model.

\subsection{Semiclassical Wave Function}

    Having found the solution to the classical problem we proceed to
    construct the semiclassical wave function as discussed by Berry
    \cite{Berry83}.
    The idea is to associate wave functions $\Psi(\phi)$ to N-dimensional
    Lagrangian surfaces $\Sigma$, in the phase space $(\phi$, $\pi_\phi)$
    (in our case, N=1). The association between $\Sigma$ and $\Psi$ is
    purely geometric, the dynamics of the problem will be introduced later,
    when we evolve the surface. We take the initial surface $\Sigma_o$ as
    the invariant curve of the classical Hamiltonian for a massless field
    (m=0).

    \be h^2={\pi_ \phi}^2+\phi^2={E_o}^2 \ee

    Where $E_o=2\omega_o(k + 1)$, $\omega_o=1$, and $k$ is a natural
    number, as in Sommerfeld quantization rules.

    Our task is to associate $\Sigma$ to a wave function

    \be \Psi(\phi)=A(\phi)\exp (iS(\phi)) \ee

    Where \cite{VanVleck28}

    \be A^2(\phi)=K|{\delta^2S\over \delta\phi\,\delta I}|=K|{d\varphi\over
    d\phi^o}|=K |{\omega_o\over \pi^o}| \ee

    with $K$ a constant of proportionality, and $S$ is a solution of the
    classical Hamilton Jacobi equation, parametrized by $I$.

    If we now generalize this to a curve $\Sigma_n$, obtained by
    evolving $\Sigma_o$ n times through the classical map of equations
    (\ref{a}) and (\ref{b}), the probability density associated
    to $\Sigma_n$ is given by

    \be {A_n}^2=|{d\varphi\over  d\phi^n}|=|{d\varphi\over  d\phi^o}|\,
    |{d\phi^o\over  d\phi^n}|={A_o}^2|{d\phi^o\over  d\phi^n}| \ee

    To obtain the phase of the associated wave function we proceed as
    follows. The phase difference between two
    points separated $\delta\phi$ on the surface $\Sigma$ is

     \bea
     \Delta S_o = S_o(\phi^o,I)-S_o(0,I)&=&
     \int\limits_0^{\phi^o}\pi^od\phi^o=
     \int\limits_0^\varphi 2I\cos^2(\varphi)d\varphi \nonumber\\
     &=& I\varphi+{I\over  2\sin(2\varphi)} \eea

     And the phase difference between two points on the curve $\Sigma_n$
     (fig.\ref{Fig2}), reads

     \be \Delta S_n=\Delta S_{n-1}+{1\over 2}({\pi_\phi}^n\phi^n -
     {\pi_\phi}^{n-1}\phi^{n-1}-E_n\delta a)  \ee

     Where we neglected a global phase $\sigma_n(\phi^n(0),I)$.

     Since the function $\pi^n(\phi)$ is multivalued, the wave function
     corresponding to the curve $\Sigma_n$ is given by the superposition
     principle as

     \be \Psi^n(\phi)=\sum_\alpha {A_\alpha}^n(\phi) \, \exp [{i\over \hbar}
     {S_n}^\alpha(\phi,I)+{\pi\over 2}\mu] \ee

     where $\alpha$ labels the different branches of $\Sigma_n$ for a given
     value of $\phi$, and $\mu$ is the Maslov index associated to each fold
     of the curve $\Sigma_n$ \cite{MaslovFedorink81,Arnold78}.

\subsection{Solution to the Wheeler-DeWitt Equation}

     Following the canonical quantization procedure, we obtain the
     Wheeler-DeWitt equation for this model:

     \be {1\over  2}[{\delta^2\over \delta a^2}-{\delta^2\over \delta\phi^2}+
     {\omega_n}^2(a)\phi^2]\Psi(a,\phi)=0 \ee

     where $\omega_n(a)$ is defined in eq.(\ref{a}) and we have chosen the
     factor ordering so that the term in second derivatives becomes the
     Laplacian operator in the minisuperspace metric \cite{Keifer88}.
     Within the range $n\delta a \leq a< (n+1) \delta a$, we may expand

     \be \Psi_n(a,\phi)=\sum_j^\infty [{A_n}^j {F_n}^{j^+}(a,\phi)+
     {B_n}^j {F_n}^{j^-}(a,\phi)] \label{d} \ee

     where

     \be {F_n}^{j^\pm}(a,\phi)={1\over \sqrt{2} \root 4 \of {{E_n}^j}}
     \, \exp[\mp i \sqrt{{E_n}^j}a]{\Phi_n}^j(\phi) \ee

     and

     \be {\Phi_n}^j(\phi)=({\omega_n\over \pi})^{1/4} {1\over  (2^j j!)^{1/2}}
     \, \exp [-{1\over 2}\omega_n\phi^2] h^j (\sqrt{\omega_n}\phi ) \ee

     where $h_j(x)$ is the Hermite polynomial of grade $j$
     \cite{AbramowitzStegun64} and ${E_n}^j=\omega_n (2j+1)$.

     Asking for continuity in the wave function and its normal
     derivative at the border of each step, we obtain the recurrence
     formulae for the coefficients of the expansion

     \be {A_{n+1}}^i=\sum_j^\infty [{A_n}^j ({F_{n+1}}^{i^+},{F_n}^{j^+})+
     {B_n}^j ({F_{n+1}}^{i^+},{F_n}^{j^-})] \ee

     \be {B_{n+1}}^i=\sum_j^\infty [{A_n}^j ({F_{n+1}}^{i^-},{F_n}^{j^+})+
     {B_n}^j ({F_{n+1}}^{i^-},{F_n}^{j^-})] \label{e} \ee

     Where $(g,f)$ is the Klein Gordon inner product

     \be (g,f)=i \int d\phi \, ({g^* {\delta f\over \delta a}-
     f{\delta g^*\over \delta a}}) \label{f} \ee

     From the equations (\ref{d}) to (\ref{f}) we finally obtain
     $\Psi_n(a,\phi)$ at the border of each step,
     from which we calculate $|\Psi_n(a^*,\phi)|^2$.
     This squared amplitude will be compared to that of the Semiclassical
     wave function. Also, we obtain the Klein Gordon charge, defined by the
     scalar product of eq.(\ref{f}) as $Q_n=(\Psi_n(a,\phi),
     \Psi_n(a,\phi))$. This charge is conserved by the Wheeler-DeWitt
     equation. We shall present the details of this recurrence relations in
     Appendix A.

\section{Results}

     Figures \ref{Fig3}b and \ref{Fig5}b show unfolding of the curve
     $\Sigma$ evolved from an initial curve $\Sigma_0$
     through the classical map of eqs.(\ref{a}) and (\ref{b}).
     The most outstanding feature is the development of spiral structures
     or ``whorls'', as Berry calls them \cite{Berry83}.
     These are associated with invariant curves around a stable fixed point
     of $h$. They can arise, for example, in the twist map \cite{Meiss92},
     provided the angular frequency depends on the radius. In this case,
     points at different radii rotate around the
     central fixed point at different rates. Therefore radii map to
     spirals, and parts of $\Sigma$ traveling close to stable fixed
     points will wrap around them, as it can be seen in figures
     \ref{Fig3}b and \ref{Fig5}b.
     For these two figures (\ref{Fig3} and \ref{Fig5}) we chose two
     different values of $\delta a$, $\delta a=0.5$ and $\delta
     a=0.25$ respectively. By the 10th step the "spiral galaxy" structure
     is clearly visible in the \ref{Fig3}b curves. For the curves of
     figure \ref{Fig5}b, the spiral galaxy structure is visible
     already at the 20th
     step (it must be noted that $n=10$ and $n=20$ correspond to the
     same ``time'' of evolution for the different figures), in spite of
     having a smoother potential.

     In figures \ref{Fig3}a and \ref{Fig5}a we can appreciate the
     semiclassical wave functions associated to the phase curves
     in figures \ref{Fig3}b and \ref{Fig5}b
     respectively. The most striking features are the caustic spikes,
     which proliferate as $n$ increases and the classical curves curl
     over.

     The graphs of $|\Psi|^2$, obtained from the exact solutions of
     the Wheeler-DeWitt equation, are shown in figures \ref{Fig4}b and
     \ref{Fig6}b for two different
     values of $\delta a$, $\delta a =0.5$ and $\delta a =0.25$
     respectively. To study the corresponding quantum map
     (i.e. eq. \ref{d})
     we must choose as initial state $\Psi_0$ a wave function associated
     to $\Sigma_0$. Because $\Sigma_0$ is an invariant curve of $h_0$,
     $\Psi_0$ must be an eigenstate of this operator. For the curves mapped
     on figures \ref{Fig4}b and \ref{Fig6}b, $\Psi_0$ was chosen to
     be the 5th eigenstate
     of $h_0$. The graph of $|\Psi|^2$ for $n=0$ clearly shows the
     association with the initial curve $\Sigma_0$, with maxima at the
     caustics of the projection of $\Sigma$, and, between these,
     near harmonic oscillations. These oscillations correspond to the
     interference of two waves, associated with the intersection of
     $\pi_i(\phi)$ with the fiber $\phi=$ constant, which make
     up the WKB function of figs. \ref{Fig3}a and \ref{Fig5}a.

     By the 10th iteration (20th for $\delta a=0.25$) the WKB
     $|\Psi|^2$  (figs. \ref{Fig3}a (resp. \ref{Fig5}a))
     has developed a complexity which cannot be observed in
     the real $|\Psi|^2$. The caustics are evidently
     related to features of the wave function  until $n=10$ (resp. $n=20$)
     but thereafter there is no obvious association. This is related
     to the fact that for large $n$ the neighboring caustics in
     fig.\ref{Fig3}a (resp. fig.\ref{Fig5}a) are closer than the de Broglie
     wavelength, so they cannot
     be associated with features of $\Psi$ \cite{Berry83}. This suggests,
     in the spirit of the smoothing procedure, that a better match between
     classical and quantal calculations will be obtained if we smooth,
     in some sense, the graph of $|\Psi|^2$. Decoherence, or some other
     process that can eliminate details of the WKB function, could provide
     a suitable smoothing mechanism.

     In our numerical calculation of the exact wave function, we have
     computed only the first 100 terms of the defining series eq.(\ref{d}).
     As a check that this truncation does not impair the accuracy of
     our results, we show in figs. \ref{Fig4}a and \ref{Fig6}a a
     logarithmic plot of
     these coefficients. It is clearly seen from these plots that the
     coefficients decay exponentially, thus ensuring that the tail of
     the series does not influence the value of the wave function within
     the accuracy of our calculations. Observe that the decay of the
     expansion coefficients for $\delta a=0.25$ (fig.\ref{Fig6}a)
     is markedly
     faster than for $\delta a=0.5$ (fig.\ref{Fig6}a), as
     we should expect for
     a smoother evolution. As another check of the numerical accuracy,
     we computed the Klein Gordon charge for each step, verifying that
     it was conserved.

\section{Discussion}

     In this paper we have presented a simple cosmological model where
     the WKB approximation fails to reproduce the exact solution to the
     Wheeler-DeWitt equation for large Universes. The breakdown of the WKB
     approximation follows the same pattern than in semiclassical physics
     of chaotic systems, and it is associated to the development of small
     scale structure in the wave function. This result puts in doubt the
     so-called ``WKB interpretacion of Quantum Cosmology'', at least in its
     original formulation \cite{Vilenkin89}. More complex models may provide
     a smoothing mechanism for the wave function, thus restoring the WKB
     approximation.

     In this model, the breakdown of the WKB approximation
     follows from the joint action of two effects, namely, the twist caused
     by the dynamics within each
     step, and the ``particle creation'' effect, that produces the excitation
     of higher eigenstates at each threshold. Since each of these effects
     are present in the continuum limit, $\delta a \to 0$, we may conclude
     that the WKB approximation will not be restored there. This is put
     into evidence by the fact that halving $\delta a$ leads to a
     stronger, rather than weaker, faillure of the WKB aproximation.

\acknowledgments

     We wish to thank Claudio El Hasi for his very kind help.
     This work was partially supported by the Consejo Nacional de
     Investigaciones Cient\'{\i}ficas y Te\'cnicas (Argentina);
     Universidad de Buenos Aires; and by Fundaci\'on Antorchas.

\appendix
\section{Recurrence Formulae for the Exact Wave Function}

     The scalar products in equations (\ref{e}) and (\ref{f}) may be
     computed from the recurrence relations

       \be ({F_{n+1}}^{i+},{F_n}^{j+}) = {[\sqrt {{E_{n+1}}^i}
       + \sqrt{{E_n}^j}] \over 2 ({E_{n+1}}^i \, {E_n}^j)^{1/4}} \exp
       [i ( \sqrt {{E_{n+1}}^i} - \sqrt {{E_n}^j}) a^* ]
       \langle n+1;i | n;j \rangle \ee

       \be ({F_{n+1}}^{i+},{F_n}^{j-}) = {[\sqrt {{E_{n+1}}^i}
       - \sqrt{{E_n}^j}] \over 2 ({E_{n+1}}^i \, {E_n}^j)^{1/4}} \exp
       [i ( \sqrt {{E_{n+1}}^i} + \sqrt {{E_n}^j}) a^* ]
       \langle n+1;i | n;j \rangle \ee

       \be ({F_{n+1}}^{i-},{F_n}^{j+}) = {[- \sqrt {{E_{n+1}}^i}
       + \sqrt{{E_n}^j}] \over 2 ({E_{n+1}}^i \, {E_n}^j)^{1/4}} \exp
       [i (- \sqrt {{E_{n+1}}^i} - \sqrt {{E_n}^j}) a^* ]
       \langle n+1;i | n;j \rangle \ee

       \be ({F_{n+1}}^{i-},{F_n}^{j-}) = {[- \sqrt {{E_{n+1}}^i}
       - \sqrt{{E_n}^j}] \over 2 ({E_{n+1}}^i \, {E_n}^j)^{1/4}} \exp
       [i (- \sqrt {{E_{n+1}}^i} + \sqrt {{E_n}^j}) a^* ]
       \langle n+1;i | n;j \rangle \ee

  In these equations,

  \be \langle n+1;i | n;j \rangle =  \int dx \, {{\Phi_{n+1}}^i}^* (x)
  \, {\Phi_n}^j (x) \ee

  These brackets can be written in recursive form

  \be \langle n+1;j | n;i \rangle = {\beta_n \over \alpha_n}
  \sqrt {j-1 \over j} \, \langle n+1;j-2 | n;i \rangle + {1 \over \alpha_n}
  \sqrt {i \over j} \langle n+1;j-1 | n;i-1 \rangle \ee

  for  $i \leq j$, and

   \be \langle n+1;j | n;i \rangle = -{\beta_n \over \alpha_n}
   \sqrt {i-1 \over i} \, \langle n+1;j | n;i-2 \rangle + {1 \over \alpha_n}
   \sqrt {j \over i} \, \langle n+1;j-1 | n;i-1 \rangle \ee

  if $j \leq i$, where \cite{Parker69}

  \be \langle n+1;0 | n;0 \rangle = \sqrt {1 \over \alpha_n} \ee

  and

   \be \alpha_n = {1 \over 2} [\sqrt {\omega_{n+1} \over \omega_n} +
        \sqrt {\omega_n \over \omega_{n+1}}] \ee

    \be \beta_n = {1 \over 2} [\sqrt {\omega_{n+1} \over \omega_n} -
        \sqrt {\omega_n \over \omega_{n+1}}] \ee

$\omega_n$ is the step frequency given in eq.(\ref{c})

\begin{figure}
\caption{Comparison of the potential $V(a)=a^2$ with the potencial of the
equation (\protect \ref{v}).\label{Fig1}}
\end{figure}

\begin{figure}
\caption{The diference of phase between two points on the curve $\Sigma_n$
may be computed as the diference of phase on the curve $\Sigma_{n-1}$ plus
the phase gain from evolutin in $\delta a$\label{Fig2}}
\end{figure}

\begin{figure}
\caption{(a) Semiclassical Wave Function associated with the evolution of
the curve $\Sigma$ for a value of $\delta a=0.5$. (b) Evolution of the
curve $\Sigma$ through the classical map, defined in equations (\protect
\ref{a}) and (\protect \ref{b}), for a value of $\delta a=0.5$.\label{Fig3}}
\end{figure}

\begin{figure}
\caption{(a) Weight of the coefficients of the defining series of equation
(\protect \ref{d}) for a value of $\delta a=0.5$. (b) Evolution of the exact
Wave Function of the Universe through the quantum map, defined in equation
(\protect \ref{d}), for a value of $\delta a=0.5$.\label{Fig4}}
\end{figure}

\begin{figure}
\caption{(a) Semiclassical Wave Function associated with the evolution of
the curve $\Sigma$ for a value of $\delta a=0.25$. (b) Evolution of the
curve $\Sigma$ through the classical map, defined in equations (\protect
\ref{a}) and (\protect \ref{b}), for a value of $\delta a=0.25$.
\label{Fig5}}
\end{figure}

\begin{figure}
\caption{(a) Weight of the coefficients of the defining series of equation
(\protect \ref{d}) for a value of $\delta a=0.25$. (b) Evolution of the exact
Wave Function of the Universe through the quantum map, defined in equation
(\protect \ref{d}), for a value of $\delta a=0.25$. \label{Fig6}}
\end{figure}


\begin{references}

     \bibitem{Halliwell90}J. J. Halliwell, in Proceedings of the Jerusalem
     Winter School on Quantum Cosmology and Baby Universes, ed. T. Piran.
     Introductory Lectures on Quantum Cosmology, (1990).

     \bibitem{Kuchar92}K. V. Kuchar, in Proceedings of the 4th Canadian
     Conference on General Relativity and Relativistic Astrophysics, eds.
     G. Kunstatter, D. Vincent, J. Willimams. (World Scientific), Time
     and Interpretatons of Quantum Gravity, (1992).

     \bibitem{Vilenkin89}A. Vilenkin, Interpretation of the Wave Function of
     the Universe, Phys. Rev. {\bf D39,} 1116 (1989).

     \bibitem{Hartle86}J. B. Hartle, in Gravitation in Astrophysics,
     (Cargese 1986), eds. B. Carter, J. Hartle. (Plenum, New York),
     Predictions and Observations in Quantum Coamology, (1986).

     \bibitem{PazSinha91}J. P. Paz, S. Sinha, Decoherence and Back
     Reaction:The Origin of the Semiclassical Einstein Equations,
     Phys. Rev. {\bf D44,} 1038 (1991).

     \bibitem{Berry83}M. V. Berry, in Chaotic Behavior of Deterministic
     Systems. (Les Houches, Session 36), eds. G. Ioos, R. H. G. Hellemman
     and R. Stora. (Amsterdam: North Holland), 453 (1983).

     \bibitem{Ozorio90}A. M. Ozorio de Almeida, Hamiltonian Systems Chaos
     and Quantization, (Cambridge University Press), (1990).

     \bibitem{CalzettaElHasi93}E. Calzetta, C. El Hasi, Chaotic Friedmann-
     Robertson-Walker Cosmology, Class. Quantum Grav., {\bf 10,}
     1825 (1993).

     \bibitem{VanVleck28}J. H. Van Vleck, The Correspondence Principle in the
     Statical Interpretation of Quantum Mechanics, Proc. Natl. Acad. Sci. USA
     {\bf 14,} 178 (1928).

     \bibitem{MaslovFedorink81}V. P. Maslov, M. V. Fedorink, Semi-Classical
     Approximation in Quantum Mechanics, (Dordrecht: Reidel), (1988).

     \bibitem{Arnold78}V. I. Arnol'd, Mathematical Methods of Classical
     Mechanics. New York:Springer, (1978).

     \bibitem{Keifer88}C. Keifer, Wave Packets in Minisuperspace, Phys. Rev.
     {\bf D38,} 1761 (1988).

     \bibitem{AbramowitzStegun64}M. Abramowitz, I. A. Stegun, Handbook
     of Mathematical Functions. Washington DC: U.S. National Bureau of
     Standards, (1964).

     \bibitem{Meiss92}J. D. Meiss, Symplectic Maps, Variational Principles,
     and Transport, Rev. Mod. Phys. {\bf 64}, 1057 (1992).

     \bibitem{Parker69}L. Parker, Phys. Rev. {\bf 183}, 1057 (1969).

\end{references}
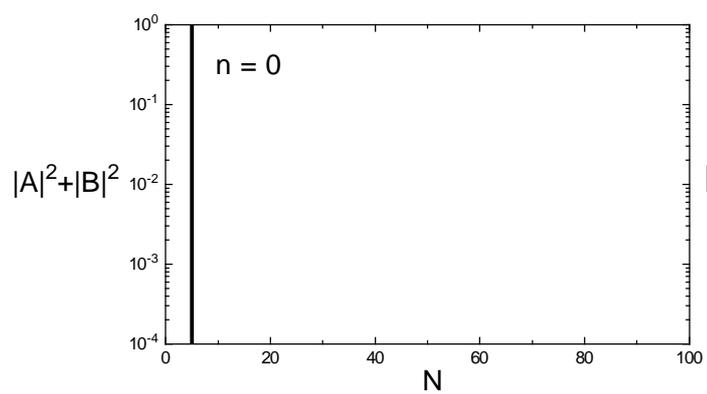
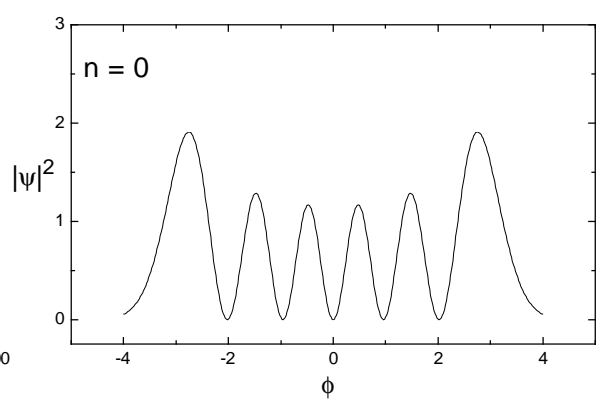
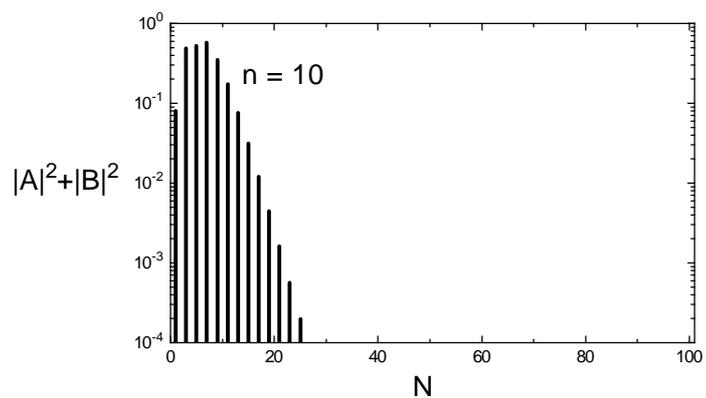
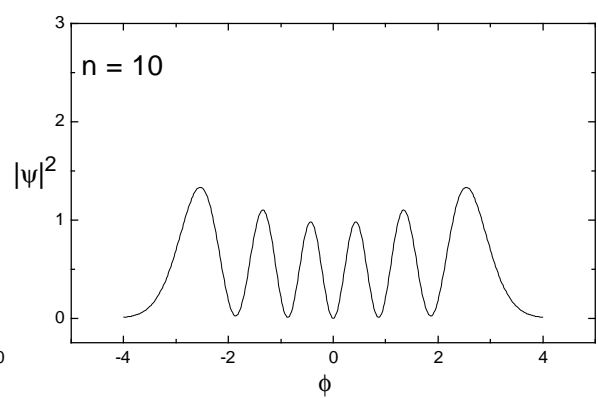
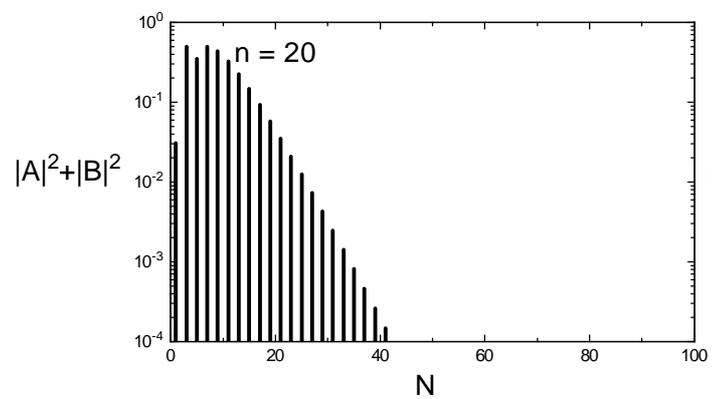
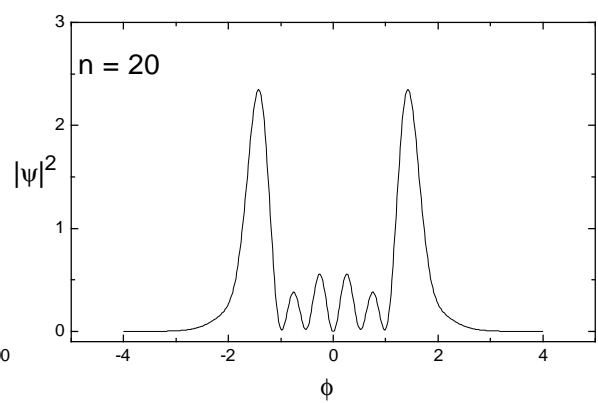
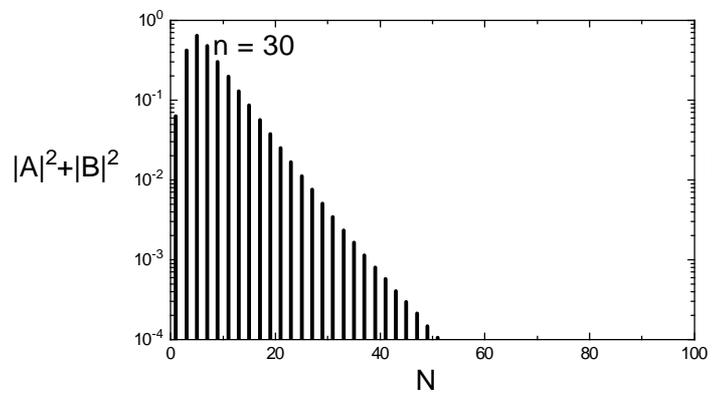
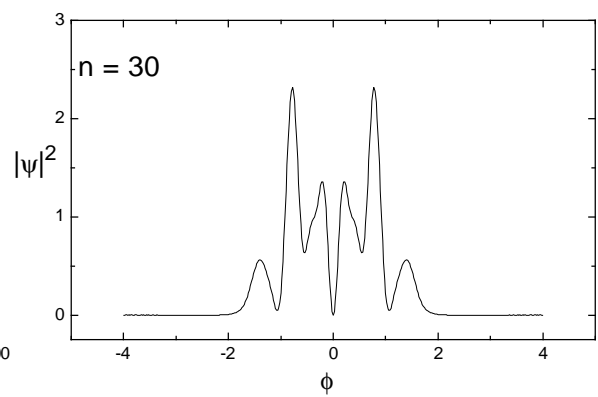
\end{document}